\documentclass{article}
\usepackage[T1]{fontenc}
\usepackage[utf8]{inputenc}
\usepackage{ismir}
\usepackage{amsmath,amssymb,cite,url}
\usepackage{graphicx}
\usepackage{subcaption}
\usepackage{color}
\usepackage{booktabs}
\usepackage{multirow}
\usepackage{bm}

\ifdefined
\else
  
\fi

\title{Beyond Dry References: Learning Relative Audio Effects\\Representations via Contrastive Distance Learning}

\multauthor
  {Xinlu Liu$^{1}$ \quad Huibin Lin$^{1}$ \quad Weixing Wei$^{2}$ \quad Zhenhai Yan$^{1}$}
  {$^{1}$ Galaxy Audio Effect Team, Tencent Music Entertainment \quad $^{2}$ Independent Researcher \\
   {\small\texttt{\{fayelliu, binglin, zhenhaiyan\}@tencent.com, wwxzju@gmail.com}}}

\def\authorname{X. Liu, H. Lin, W. Wei, and Z. Yan}

\begin{document}

\maketitle

\begin{abstract}
Audio effects (Fx) representation learning plays a key role in intelligent music production, including automatic mixing and Fx style transfer.
Existing methods typically rely on dry or nearly dry references for effect modeling, yet truly unprocessed audio is rarely available in practice, as real recordings inevitably reflect the microphone, room acoustics, and preceding signal processing.
Instead of pursuing absolute effect encodings, we argue that the \emph{relative effect distance} between audio signals is more meaningful for real-world music production.
Motivated by this, we propose \textbf{RelFx}, a contrastive learning framework that learns relative effect transformations from general audio collections without requiring dry references during representation training.
Our approach uses a dual-branch Siamese encoder equipped with cross-attention and differential gating fusion 
to infer the shared effect transformation from a reference clip and an effect-processed, content-related clip.
We further propose an antisymmetric fusion variant for bidirectional effect encoding, such that swapping the input order directly produces a nearly sign-reversed embedding, a property not explored in earlier work.
Moreover, our dry-reference-free formulation eliminates the reliance on dry multitrack datasets and enables training on effect-bearing audio.
Experiments on Fx style transfer demonstrate state-of-the-art performance under the standard Fx-Encoder++ MUSDB18 evaluation protocol, consistently outperforming existing approaches across all four instrument categories.
Audio examples and source code are available at \url{https://relative-fx.github.io}.
\end{abstract}

\begin{table*}[t]
    \centering
    \caption{Content similarity (cosine similarity) between 10-second audio clip pairs under different pairing conditions, measured with Fx-Encoder++ embeddings (2048-dim) ~\cite{yeh2025fxencoderpp} on 2{,}000 songs.}
    \label{tab:content_sim}
    \small
    \begin{tabular}{clcccc}
        \toprule
        \textbf{Cond.} & \textbf{Pairing Strategy} & \textbf{\#Pairs} & \textbf{Mean} & \textbf{Median} & \textbf{$\geq$0.9 (\%)} \\
        \midrule
        A & Same song, same section, adjacent clips             & 6,980   & 0.938 & 0.960 & 82.61 \\
          & \quad -- verse only                      & 1,788   & 0.956 & 0.965 & 94.57 \\
          & \quad -- chorus only                     & 3,015   & 0.954 & 0.967 & 91.34 \\
        \midrule
        B & Same song, same section, non-adjacent clips         & 5,042     & 0.918   & 0.945   & 69.50 \\
        C & Same song, different sections, adjacent clips         & 4,437     & 0.838   & 0.858   & 31.15 \\
        D & Same song, different sections, non-adjacent clips            & 121,466     & 0.783   & 0.803   & 17.00 \\
        E & Different songs, different sections, non-adjacent clips            & 199,991     & 0.684   & 0.694   & 0.23 \\
        \bottomrule
    \end{tabular}
\end{table*}

\section{Introduction}\label{sec:introduction}

Audio effects (Fx) such as equalization, compression, reverb, and delay are fundamental tools in music production.
With the development of intelligent music production systems~\cite{steinmetz2022automix, martinez2022fxnorm}, the computational representation and reasoning of Fx have become increasingly crucial.
Applications including automatic mixing~\cite{steinmetz2022automix, martinez2021deep, shiscalable}, Fx style transfer~\cite{steinmetz2022style,steinmetz2024afxrep, koo2022remaster, koo2025ito} and mixing style transfer~\cite{vanka2024diffmst, koo2023music} all require robust Fx representations. 
Recent text-driven effects control and tool-guided audio post-processing~\cite{chu2025text2fx, ki2026fxsearcher, doh2025llm2fx} also rely on high-quality Fx representations.

Recent work on Fx representation learning has made significant progress~\cite{steinmetz2024afxrep, koo2023music, yeh2025fxencoderpp}.
However, these methods share a critical limitation: they aim to learn absolute Fx representations under the common assumption of access to dry or nearly dry audio, which can be difficult to obtain in practice.

This assumption is problematic for three reasons.
First, truly dry audio is practically unattainable---every recording inherently carries acoustic characteristics from the recording environment, microphone, and signal chain.
To mitigate this issue, various methods have been explored to remove existing effects from audio signals and approximate dry references~\cite{rice2023removal, martinez2022fxnorm, steinmetz2022efficient, saito23dereverberation}. For instance, prior work employs Fx-Normalization~\cite{martinez2022fxnorm}, a preprocessing step that attempts to ``normalize'' audio to a common effects baseline.
However, this normalization is inherently approximate and introduces systematic errors.
Second, the requirement for dry or nearly dry audio restricts training data to multitrack datasets such as MUSDB18~\cite{rafii2017musdb18} and MoisesDB~\cite{pereira2023moisesdb}, which contain only a limited number of songs, thereby constraining data-driven contrastive learning.
Third, the absolute, single-input paradigm only supports \emph{unidirectional} modeling (i.e., adding effects to a reference signal). It cannot represent \emph{effect reduction} such as dereverberation or effect removal, since it describes static signal states rather than relative transformations.

We propose to reformulate the problem.
In practice, mixing engineers seldom reason solely in terms of absolute effect states; instead, they primarily think in terms of \emph{relative distances}---``how do I get from the current sound to the target sound?''
Motivated by this observation, we introduce \textbf{RelFx}, a contrastive learning framework for \emph{relative Fx distance learning} using audio pairs that need not be dry or effect-normalized.
Given content-related clips $x$ and $x'$ and a known effect transformation $\theta$, our model learns to encode the transformation from the ordered pair $\bigl(x,\operatorname{Fx}(x';\theta)\bigr)$, regardless of what effects the original audio may already contain.

Our key contributions are:

\begin{enumerate}

\item We introduce a relative-distance formulation for Fx learning that removes dry-reference dependence during representation learning and broadens the usable training data beyond dry multitracks.

\item We present an effective framework that explicitly models signal transformations, overcoming limitations of traditional dry-wet supervision.

\item We achieve state-of-the-art effects style-transfer performance under the standard Fx-Encoder++ MUSDB18 evaluation protocol across four instrument categories.

\item We propose an antisymmetric bidirectional fusion module that provides direction-aware relative-effect embeddings, with input-order swapping producing an approximately sign-reversed representation.

\end{enumerate}

\section{Related Work}\label{sec:related}

\subsection{Related Applications in Intelligent Music Production}
Fx style transfer is a core task in intelligent music production, which aims to capture the timbral characteristics of a reference signal. 
Related methods include optimization with differentiable signal processing~\cite{steinmetz2022style,vanka2024diffmst,koo2025ito}, blind processor-parameter estimation~\cite{peladeau2024blind,peladeau2025audio}, and end-to-end neural networks~\cite{koo2023music}.

Automatic music mixing is another key task, focusing on balancing multi-track audio and shaping timbre to produce professional mixtures. Recent work often employs effects modeling and parameter control to imitate target mixing styles or automate the mixing pipeline~\cite{steinmetz2022automix,steinmetz2024afxrep}.

Both tasks rely on effective Fx representations. Accurate modeling of effect transformations is essential for generating high-quality results.

\subsection{Fx Representation Learning}

Learning representations of Fx has been explored from several perspectives.

AFx-Rep~\cite{steinmetz2024afxrep} learns self-supervised audio production embeddings through effect and preset classification, and is trained on clean, unprocessed dry audio instead of approximated or effect-normalized references.
Fx-Encoder~\cite{koo2023music} introduces contrastive learning for mixture-level Fx representations, while Fx-Encoder++~\cite{yeh2025fxencoderpp} extends this with an instrument-wise embedding extractor guided by CLAP~\cite{wu2023clap} queries, and relies on Fx-Normalization~\cite{martinez2022fxnorm} to approximate dry references.
General-purpose audio representations (VGGish~\cite{hershey2017vggish}, CLAP~\cite{wu2023clap}) capture broad semantic properties but lack fine granularity for effect-specific modeling.

\section{Method}\label{sec:method}

\subsection{Problem Formulation}\label{subsec:formulation}

Existing methods encode the absolute effect state of an audio signal $E(x) \rightarrow \bm{z}$, which requires a consistent effect baseline (dry audio). We instead formulate Fx representation as encoding a \emph{relative transformation}.

Given a reference clip $x$, a source clip $x'$, and an effect transformation $\theta$, we define the processed-source clip as $\tilde{x}=\operatorname{Fx}(x';\theta)$. 
Our encoder takes the ordered pair $(x,\tilde{x})$ as input:
\begin{equation}
z=E(x,\tilde{x})=E\bigl(x,\operatorname{Fx}(x';\theta)\bigr).
\label{eq:relative_encoding}
\end{equation}
Here, $x'=x$ is a waveform-aligned special case. During representation training, $x$ and $x'$ are adjacent, non-overlapping clips sampled from the same musical section, as described in Sec.~3.3.1. Neither clip is required to be dry or effect-normalized.

For observation $q$, let
$z_q(\theta)=E\bigl(x_q,\operatorname{Fx}(x'_q;\theta)\bigr)$.
The contrastive objective encourages the same transformation applied to different musical content to produce similar embeddings:
\begin{equation}
d\bigl(z_a(\theta),z_b(\theta)\bigr)
\ll
d\bigl(z_a(\theta),z_c(\theta')\bigr),
\label{eq:relative_objective}
\end{equation}
where $\theta\neq\theta'$, the observations contain different musical content, and $d(\cdot,\cdot)$ denotes a generic embedding-space distance; the concrete losses are defined in Sec.~3.3.3.

We evaluate whether the same effect transformation produces consistent embeddings across different musical content, providing evidence that the learned representation is relatively robust to content variations.

\subsection{Model Architecture}\label{subsec:architecture}

\begin{figure}[t]
    \centering
    \includegraphics[
        alt={A weight-shared two-branch CNN encodes a reference segment and a transformed segment, exchanges intermediate features through bidirectional cross-attention, pools both branches, and combines them with Diff-Gate fusion before projection.},
        width=\linewidth
    ]{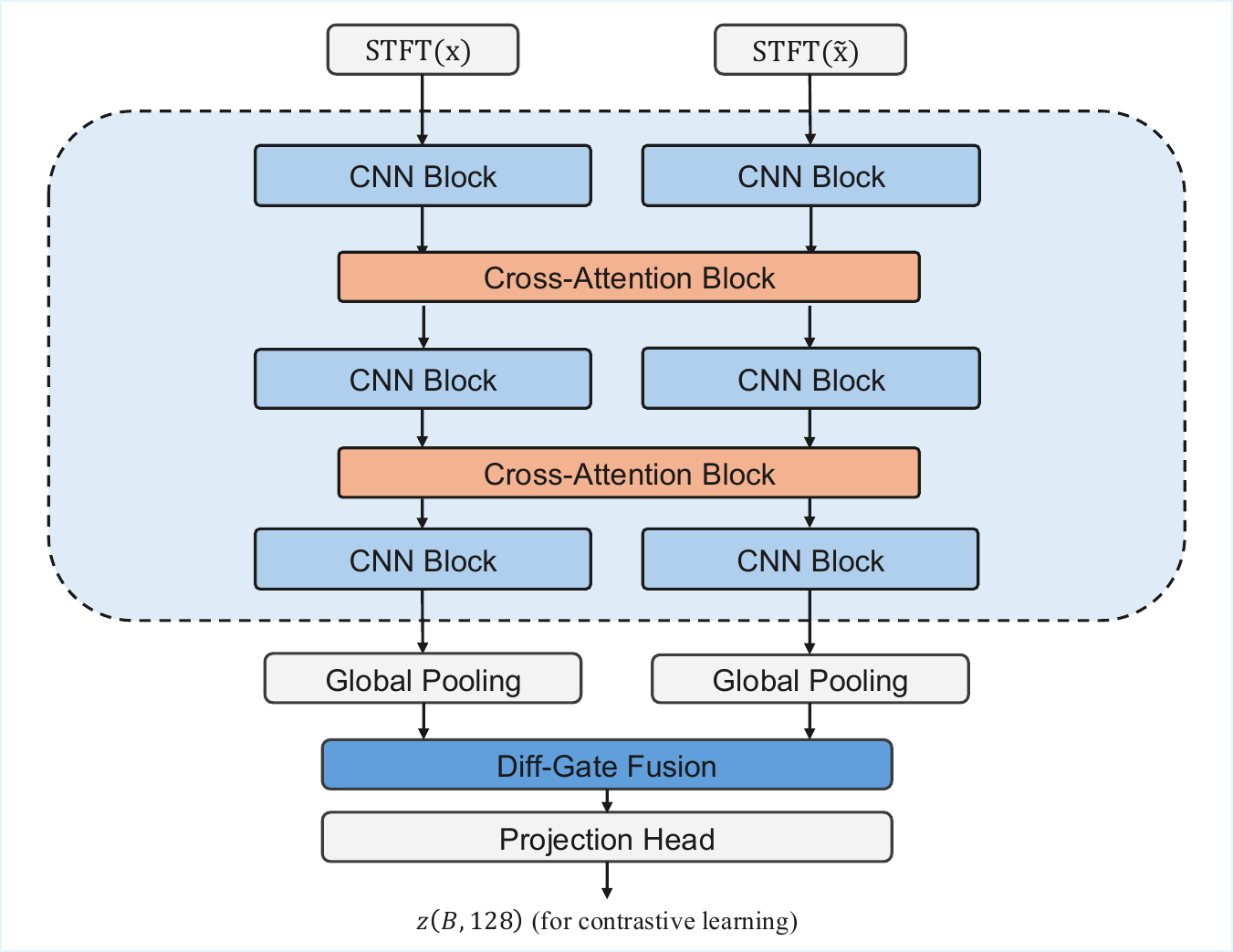}
    \caption{RelFx architecture. Weight-shared CNN branches exchange features through bidirectional cross-attention before Diff-Gate fusion. The antisymmetric variant uses the fusion in (Eqs.~\ref{eq:antisym_diffgate} -~\ref{eq:antisym_diffgate_2}) and the projection in Sec.~3.2.5.}
    \label{fig:architecture}
\end{figure}

Guided by the above relative transformation formulation, we propose a dual-branch architecture as illustrated in \figref{fig:architecture}.

\subsubsection{Spectrogram Frontend}

For protocol compatibility, we retain the Fx-Encoder++ frontend~\cite{yeh2025fxencoderpp}: both inputs are converted to log-mel spectrograms using an STFT with $n_\text{fft}=2048$ and hop size 512.
Following CNN14~\cite{kong2020panns}, we use 64 mel bins (50--18{,}000\,Hz), log-scaling, and normalization to $[-1,1]$.

\subsubsection{Siamese CNN Backbone}

We employ a CNN14-style ~\cite{kong2020panns} backbone with six convolution stages, channel dimensions $[64, 128, 256, 512, 1024, 2048]$. 
The two branches share all weights (Siamese configuration~\cite{riad18sampling}), ensuring that both signals are encoded in the same representation space.

\subsubsection{Cross-Attention Interaction}
To enable each branch to incorporate the intermediate features of the other and better model the transformation between the two signals,
we insert bidirectional cross-attention after stages 3 (256 channels) and 5 (1024 channels)
.
Each block uses 4 heads, $4\times$ spatial pooling to limit memory, and residual scale 0.1 to preserve backbone features:
\begin{equation}
    \bm{h}_\text{ref}' = \bm{h}_\text{ref} + \alpha \cdot \text{CrossAttn}(\bm{h}_\text{ref}, \bm{h}_\text{proc})
\end{equation}
where $\alpha = 0.1$. The cross-attention is applied symmetrically to both branches~\cite{Lin23AvSepformer, nihal2025cross}.

\subsubsection{Differential Gating Fusion}

After global pooling, we obtain embeddings $e_{\mathrm{ref}},\,e_{\mathrm{proc}}\in\mathbb{R}^{2048}$.
We propose two fusion variants that combine the two branches into a single effect embedding.

\textbf{Base model.}
Our base fusion mechanism uses a learned gate to balance the relative change signal with the absolute processed signal:
\begin{equation}\label{eq:diffgate}
    \bm{e}_\text{fx} = \bm{g} \odot (\bm{e}_\text{proc} - \bm{e}_\text{ref}) + (1 - \bm{g}) \odot \bm{e}_\text{proc}
\end{equation}
\begin{equation}
    \bm{g} = \sigma(\bm{W}[\bm{e}_\text{ref}; \bm{e}_\text{proc}] + \bm{b})
\end{equation}
where $\sigma$ is the sigmoid function and $[\cdot ; \cdot]$ denotes concatenation.
The differential term captures the effect-induced relative change between signals, while the absolute term retains the content and timbral characteristics of the processed signal~\cite{yu2025dgfnet}.
The gate adaptively balances these two sources depending on the effect type.

\textbf{Bidirectional-compatible variant.}
To enforce antisymmetry at the fusion stage, we revise the fusion mechanism with two key adjustments: (1) the concatenation-based gate input is replaced with element-wise addition, making the gate output invariant to input order; (2) we remove the standalone static branch term $\bm{e}_\text{proc}$ and retain only the differential interaction term:
\begin{equation}\label{eq:antisym_diffgate}
    \bm{e}_\text{fx} = \bm{g} \odot (\bm{e}_\text{proc} - \bm{e}_\text{ref})
\end{equation}
\begin{equation}\label{eq:antisym_diffgate_2}
    \bm{g} = \sigma(\bm{W}(\bm{e}_\text{ref} + \bm{e}_\text{proc}) + \bm{b})
\end{equation}
Since the gate $\bm{g}$ is order-invariant and the differential term $(\bm{e}_\text{proc} - \bm{e}_\text{ref})$ reverses its sign when swapping the two branch embeddings, 
the fusion output is antisymmetric by construction.

\textbf{Implications for bidirectional encoding.}
A practical consequence of the antisymmetric fusion is that swapping an ordered input pair yields an approximately sign-reversed relation embedding:
\begin{equation}\label{eq:bidirectional}
    E(A,B) \approx -E(B,A).
\end{equation}
For the waveform-aligned case $B=\operatorname{Fx}(A;\theta)$, this represents the reverse direction of the same input--output relation. The antisymmetric variant therefore provides a unified, direction-aware representation of relative effect changes.
In our experiments, we use the base model for the main style-transfer evaluation and separately train the antisymmetric variant to evaluate this structural property.

\subsubsection{Projection Head}

A two-layer MLP projects $\bm{e}_\text{fx}$ to a 128-dimensional embedding with L2 normalization:
$\bm{z} = \text{L2Norm}(\text{MLP}(\bm{e}_\text{fx})) \in \mathbb{R}^{128}$.
The base model uses ReLU activation.
For the bidirectional-compatible variant, we replace ReLU with $\tanh$---an odd function satisfying $\tanh(-x) = -\tanh(x)$---which helps preserve the sign-reversal property through projection. Because the linear layers include learned biases, antisymmetry of the final L2-normalized embedding is approximate and is evaluated empirically in Sec.~\ref{subsec:bidirectional}.

\begin{figure}[t]
    \centering
    \includegraphics[
        alt={Four observations formed from different audio segments. Observations sharing the same effect transformation form positive pairs, while observations with independently sampled transformations form negative pairs.},
        width=0.8\linewidth
    ]{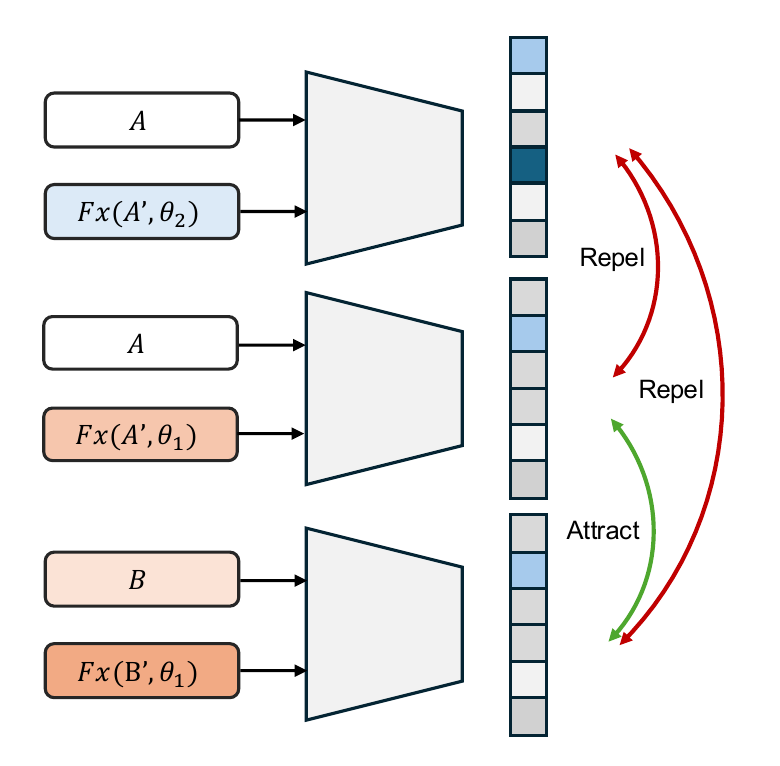}
    \caption{Illustration of our contrastive learning framework. Positive pairs share the same effects transformation $\theta$ applied to different audio content, while negative pairs use independently sampled configurations. Cross-segment sampling ensures content-effect disentanglement.}
    \label{fig:contrastive}
\end{figure}

\subsection{Training}\label{subsec:training}

\subsubsection{Data Preparation}

\figref{fig:contrastive} summarizes the training-pair construction.
For each training item, we sample two distinct audio tracks indexed by $q\in\{a,b\}$, a pair of non-overlapping clips $(x_q,x'_q)$ from each track, and one effect configuration $\theta$.
We encode $\bm{z}_q(\theta)=E(x_q,\text{Fx}(x'_q,\theta))$; $(\bm{z}_a(\theta),\bm{z}_b(\theta))$ is positive, whereas negatives use a different configuration $\theta'$.
For full-mix audio, the two tracks come from different songs; for MoisesDB, they may also be different stems from the same song.

\textbf{Cross-segment sampling.}
Within an observation, $(x_q,x'_q)$ are adjacent, non-overlapping clips from one structural section (e.g., verse), retaining related musical context without sample-wise waveform alignment.
This within-observation pairing discourages content-based shortcuts, whereas positive observations are formed from independently sampled tracks.

\textbf{Segment-level content analysis.}
Table~\ref{tab:content_sim} provides a content-similarity analysis supporting this sampling choice.
For Fx-Encoder++ embeddings~\cite{yeh2025fxencoderpp} over 2{,}000 songs, mean similarity falls from $\geq0.918$ for same-section pairs (A/B), to 0.783--0.838 across sections (C/D), and 0.684 across songs (E).
These results support using nearby clips from the same song section, since they contain some waveform variation while still sharing similar musical content. RelFx is also evaluated on song-disjoint splits and MUSDB18. A systematic evaluation of its robustness across different instruments is left for future work.

\subsubsection{Effects Chain}

Following Fx-Encoder++~\cite{yeh2025fxencoderpp}, we use a chain of 8 sequential effects (equalizer, distortion, multiband compressor, gain, stereo imager, limiter, delay, reverb) ~\cite{steinmetz2022style}, resulting in 72 continuous parameters and 8 binary activation switches. The effect order is randomly shuffled per batch; each effect is activated with a dynamically adjusted probability during training (detailed below).

\textbf{Dynamic Fx probability scheduling.}
Instead of fixed activation probabilities, we adaptively adjust each effect's activation probability based on its running training loss. At the end of each epoch, we compute the mean parameter regression loss for each effect over the past epoch, then renormalize probabilities such that effects with higher residual loss are sampled more frequently in the next epoch. All probabilities are clamped to $[p_\text{min}, p_\text{max}] = [0.3, 0.8]$ to maintain diversity and avoid single-effect dominance. Initialized with uniform probabilities, this strategy is updated throughout training, and its contribution is analyzed in Sec.~\ref{subsec:ablation}.

\subsubsection{Loss Function}

Our combined loss consists of three components:
\begin{equation}\label{eq:loss}
    \mathcal{L} = \mathcal{L}_\text{ctr} + \lambda_t \mathcal{L}_\text{triplet} + \lambda_p \mathcal{L}_\text{param}
\end{equation}

\textbf{Contrastive loss.}
For $N$ positive pairs, let $p(i)$ denote the index of the positive counterpart of $z_i$ among the $2N$ embeddings, and define
$s_{ik}=\operatorname{sim}(z_i,z_k)/\tau$ using cosine
similarity and $\tau=0.15$. We use the following modified NT-Xent objective~\cite{chen2020simclr}, where the positive logit is included twice in the denominator:
\begin{equation}
\mathcal{L}_{\mathrm{ctr}}=-\frac{1}{2N}\sum_{i=1}^{2N}
\log\frac{e^{s_{i,p(i)}}}
{2e^{s_{i,p(i)}}+
\sum_{\substack{k=1\\k\notin\{i,p(i)\}}}^{2N}e^{s_{ik}}}.
\label{eq:modified_ntxent}
\end{equation}

\textbf{Triplet loss}~\cite{liu2022selfsupervise} provides explicit margin-based separation with margin $m=0.3$:
\begin{equation}
    \mathcal{L}_\text{triplet} = \frac{1}{N}\sum_{i=1}^{N}\max(0, d(\bm{z}_i^a, \bm{z}_i^b) - d(\bm{z}_i^a, \bm{z}_i^\text{neg}) + m)
\end{equation}
where $d(\cdot,\cdot)$ denotes cosine distance, $z_i^a$ and $z_i^b$ form the $i$-th positive pair, and $z_i^{\mathrm{neg}}=E\bigl(x_i,\operatorname{Fx}(x'_i;\theta'_i)\bigr)$ reuses the anchor clips with $\theta'_i\neq\theta_i$. The triplet term enforces a margin against this content-matched negative. While the contrastive term already uses all in-batch negatives, the triplet term provides an \emph{explicit} margin against this controlled hard negative, enforcing a minimum required embedding margin between content-matched observations with distinct effect parameters.

\textbf{Parameter regression loss} enforces continuous
structure in the embedding space. Let $\bm{p}$ denote the
72-dimensional vector of continuous effect parameters and
$\bm{a}$ the eight activation switches. An auxiliary head
predicts both quantities from the embedding:
\begin{equation}
    \mathcal{L}_{\mathrm{param}}
    =
    \operatorname{MSE}_{\mathrm{masked}}
    \bigl(\hat{\bm{p}}, \bm{p}\bigr)
    +
    \lambda_a
    \operatorname{BCE}
    \bigl(\hat{\bm{a}}, \bm{a}\bigr).
\end{equation}
The masked MSE is computed only over parameters associatedwith activated effects.

We set ($\lambda_t=0.5$), ($\lambda_p=0.5$), and ($\lambda_a=0.3$), and use ($\tau=0.15$) and ($m=0.3$). All hyper parameters are fixed across ablations.

\section{Experiments}\label{sec:experiments}

\subsection{Experimental Setup}

\begin{figure}[t]
    \centering
    \begin{subfigure}[t]{0.44\linewidth}
        \centering
        \includegraphics[
            alt={Single-input effects style-transfer pipeline in which Diff-DSP produces candidate A-hat and independent encoders compare it with reference B.},
            width=\linewidth
        ]{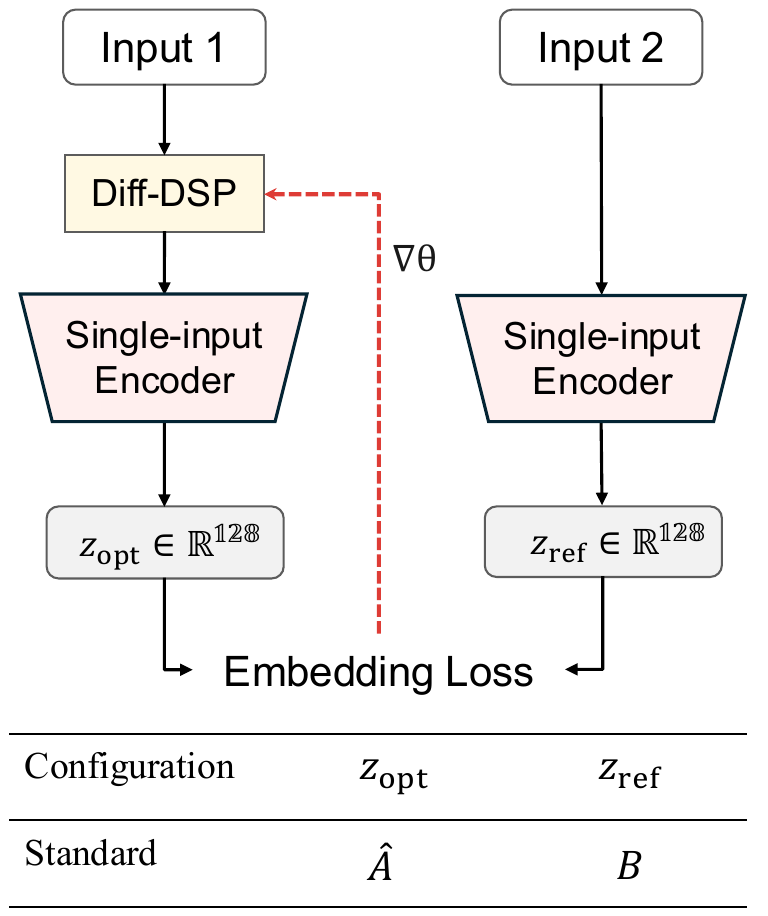}
        \caption{single-input }
    \end{subfigure}
    \hfill
    \begin{subfigure}[t]{0.54\linewidth}
        \centering
        \includegraphics[
            alt={RelFx effects style-transfer pipeline with two dual-input encoder branches and a table listing the ordered pairs for Self-ref, Standard, and Oracle configurations.},
            width=\linewidth
        ]{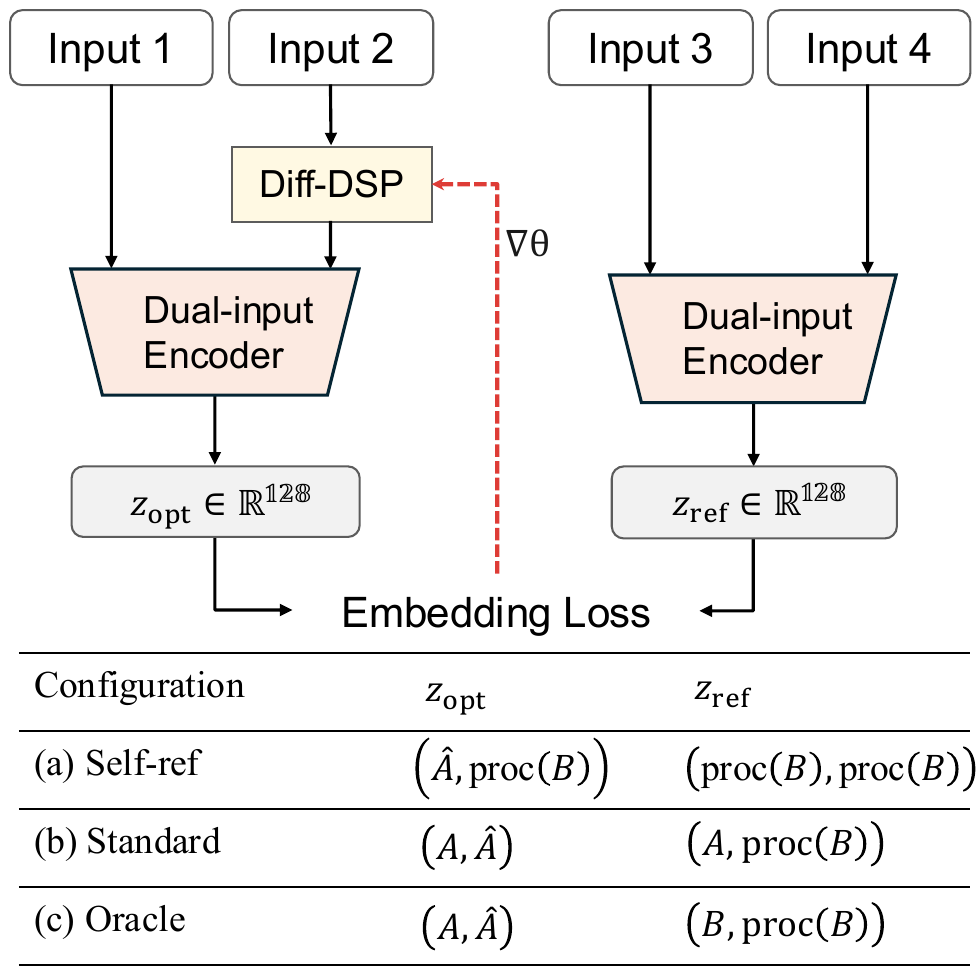}
        \caption{RelFx (dual-input)}
    \end{subfigure}
    \caption{ITO pipelines for Fx style transfer. 
Both paradigms optimize the effect-chain parameters ${\theta}$ of a differentiable DSP using gradients $\nabla{\theta}$ derived from an embedding loss.
}
    \label{fig:style_transfer}
\end{figure}

\subsubsection{Training Data}

We train on audio from two sources: (1)~a licensed internal collection of 6{,}447 mixed tracks (2{,}681 songs, 185.9\,h), and (2)~2{,}585 stems from MoisesDB~\cite{pereira2023moisesdb} (240 songs, 156.4\,h).
During training, 10-second stereo clips are sampled at 44.1~kHz, matching the sampling rate used by the DSP chain and the evaluation protocol; 
full-mix clips come from verse and chorus sections identified by music structure analysis.
The training and validation sets are split by song ID (85\% / 15\%) with a fixed split seed of 42.
For MoisesDB, we apply content density filtering (requiring $\geq$70\% non-silent frames) to exclude sparse regions during clip sampling.

\subsubsection{Training Details}

We train for 200 epochs on $4\times$ NVIDIA L20 GPUs, batch size 48 per GPU and gradient accumulation of 4 steps (effective batch size 768).
We use AdamW~\cite{loshchilov2019adamw} with learning rate $5 \times 10^{-4}$, weight decay $10^{-5}$, 10-epoch linear warmup, and cosine annealing to $10^{-6}$.

\subsubsection{Baselines}

We compare against two categories of methods: effect-specific representation models including Fx-Encoder++~\cite{yeh2025fxencoderpp} and AFx-Rep~\cite{steinmetz2024afxrep}, and general-purpose audio representation models including VGGish~\cite{hershey2017vggish} and CLAP~\cite{wu2023clap}. All baselines are evaluated using their official models and standard protocols.

\subsubsection{Evaluation Task}\label{subsubsec:evaltask}
\noindent\textbf{Fx style transfer.}
Our primary task is embedding-guided Fx style transfer, following the settings of Fx-Encoder++~\cite{yeh2025fxencoderpp}.
Given a reference audio with target effects, we use the encoder's embedding to guide gradient-based optimization of a differentiable effect chain, recovering the reference's processing from a clean source signal.
To avoid local optima, each sample is optimized with 7 random restarts using Adam (learning rate $10^{-2}$, cosine annealing) for up to 200 iterations with early stopping (patience 50), and the restart with the lowest $L_d$ is retained.
We report the median multi-resolution STFT loss $L_d$ (lower is better) between the optimized output and the ground-truth target on MUSDB18~\cite{rafii2017musdb18}, using the same 100 samples per instrument for every method.
These samples are processed with Fx-Normalization~\cite{martinez2022fxnorm} solely to align with the Fx-Encoder++ evaluation protocol; RelFx representation training itself does not require normalization.

\noindent\textbf{Effects retrieval (for ablation).}
We also adopt the effects retrieval protocol from Fx-Encoder++~\cite{yeh2025fxencoderpp} as a supplementary evaluation.
Given a processed query clip, we retrieve candidates with identical effects but distinct content from 500 samples, ranked by cosine similarity.
Performance is reported as Recall@$K$ ($K$=1,5,10), where a match in top-$K$ counts as correct.
We evaluate on MUSDB18~\cite{rafii2017musdb18} over four instruments with 1--8 effect chains, averaging results across settings.
This metric acts as a coarse-grained ablation indicator, while $L_d$ remains our primary metric.

\subsection{Fx Style Transfer Results}

\tabref{tab:matching} presents the Fx style transfer results, and \figref{fig:style_transfer} illustrates the ITO pipelines for both paradigms. We first detail the three input configurations for our model (used in the evaluation) before analyzing the quantitative results.

\begin{table}[t]
    \centering
\caption{Fx style-transfer results measured by median multi-resolution STFT loss $L_d$ (lower is better). \emph{Bi-dir.} is the antisymmetric variant (Eq.~\ref{eq:antisym_diffgate}, Sec.~\ref{subsec:bidirectional}); other ``RelFx'' rows use the base model (Eq.~\ref{eq:diffgate}). Best among non-oracle-reference settings in \textbf{bold}. The $^\dagger$\emph{Oracle-reference} condition uses ground-truth reference pairs and is not directly comparable.}
    \label{tab:matching}
    \footnotesize
    \setlength{\tabcolsep}{5.5pt}
    \begin{tabular}{lccccc}
        \toprule
        \textbf{Model} & \textbf{Drums} & \textbf{Bass} & \textbf{Vocals} & \textbf{Other} & \textbf{Avg} \\
        \midrule
        CLAP~\cite{wu2023clap}             & 1.730 & 1.390 & 1.792 & 1.662 & 1.644 \\
        VGGish~\cite{hershey2017vggish}    & 1.919 & 1.617 & 1.818 & 1.780 & 1.784 \\
        AFx-Rep~\cite{steinmetz2024afxrep} & 1.551 & 1.434 & 1.589 & 1.492 & 1.517 \\
        Fx-Encoder++~\cite{yeh2025fxencoderpp} & 1.684 & 1.445 & 1.693 & 1.583 & 1.601 \\
        \midrule
        RelFx (MoisesDB)                & 1.559 & \textbf{1.202} & 1.600 & 1.436 & 1.449 \\
        RelFx (Self-ref)                     & \textbf{1.385} & 1.269 & 1.538 & 1.404 & 1.399 \\
        RelFx (Standard)                     & 1.403 & 1.211 & 1.482 & 1.458 & \textbf{1.388} \\
        RelFx (Bi-dir.)                      & 1.522 & 1.204 & \textbf{1.430} & \textbf{1.400} & 1.389 \\
        \midrule
        RelFx (Oracle)$^\dagger$      & 1.355 & 1.169 & 1.415 & 1.326 & 1.316 \\
        \bottomrule
    \end{tabular}
\end{table}

We evaluate three ITO configurations; their exact ordered pairs are shown in \figref{fig:style_transfer}:

(a)~\emph{Self-ref} forms an identity target from the processed reference and pairs the candidate output $\hat A$ with that reference;

(b)~\emph{Standard} uses the source $A$ as a shared anchor for the target and optimized embeddings;

(c)~\emph{Oracle} defines the target from the ground-truth pair $(B,\text{proc}(B))$ while retaining Standard's optimization pair; it is reported only as an oracle-reference condition.

Among non-oracle-reference configurations, Standard achieves a lower average $L_d$ than Self-ref (1.388 vs. 1.399), while the Oracle-reference condition achieves the best result at 1.316.

When comparing against baselines, general-purpose audio representations (VGGish~\cite{hershey2017vggish}, CLAP~\cite{wu2023clap}) perform worst, 
suggesting that general-purpose semantic embeddings lack the fine-grained information required for Fx style transfer.
AFx-Rep~\cite{steinmetz2024afxrep} performs competitively on drums but falls short on vocals and other sources.
Our Standard model consistently outperforms Fx-Encoder++~\cite{yeh2025fxencoderpp}, with gains up to 16.7\% (drums), 16.2\% (bass), 12.5\% (vocals), 7.9\% (other), and an average improvement of 13.3\%.
Our Standard and Self-ref models achieve better results under the same input conditions as the baselines.

Notably, when trained on MoisesDB alone, the same training data used by Fx-Encoder++, RelFx achieves an average $L_d$ of 1.449, compared to 1.601 for Fx-Encoder++, a relative improvement of 9.5\%, suggesting that the gain is not solely attributable to the additional training data.

\subsection{Ablation Study}\label{subsec:ablation}
We conduct an ablation study to quantify the contribution of each component.
We evaluate on both the Fx retrieval task (R@$K$, averaged over 4 instruments and 1--8 Fx combinations, as defined in Sec.~\ref{subsubsec:evaltask}), and the Fx style transfer task (median $L_d$, averaged over 4 instruments).
We consider the latter metric to be more indicative of real-world performance.
All ablated models are trained for 200 epochs under identical conditions, with only the specified component removed.

Results are shown in \tabref{tab:ablation}.

\begin{table}[t]
    \centering
\caption{Ablation study on MUSDB18. $R@1$, $R@5$, and $R@10$ are reported for mixture-level retrieval (higher is better), and $L_d$ is measured on Fx style transfer under the Self-ref ITO configuration (lower is better).}
    \label{tab:ablation}
    \footnotesize
    \begin{tabular}{lcccc}
        \toprule
        \textbf{Configuration} & \textbf{R@1}$\uparrow$ & \textbf{R@5}$\uparrow$ & \textbf{R@10}$\uparrow$ & $\bm{L_d}\downarrow$ \\
        \midrule
        Full model                    & 67.3 & 79.2 & 83.7 & 1.399 \\
        \quad w/o cross-attention     & 65.6 & 78.2 & 82.8 & 1.392 \\
        \quad w/o cross-seg sampling  & 69.6 & 80.2 & 83.9 & 1.876 \\
        \quad w/o param regression    & 70.4 & 81.7 & 85.5 & 1.437 \\
        \quad w/o dynamic Fx prob     & 71.7 & 79.8 & 83.3 & 1.465 \\
        Single-branch                 & 46.1 & 62.6 & 68.8 & 1.457 \\
        \bottomrule
    \end{tabular}
\end{table}

\textbf{Cross-attention} enables the two branches to exchange information at intermediate feature levels.
Removing cross-attention reduces R@1 by 1.7 percentage points (67.3 to 65.6), while slightly improving $L_d$ from 1.399 to 1.392. This suggests that cross-attention benefits coarse-grained retrieval but is not essential for the style-transfer metric.

\textbf{Cross-segment sampling} is the most critical component for generalization.
Removing it \emph{improves} retrieval R@1 by 2.3 percentage points (69.6 vs.\ 67.3) but sharply degrades style transfer from 1.399 to 1.876 ($+34\%$), because the model learns content shortcuts rather than effect transformations.
This suggests that cross-segment sampling helps reduce content-based shortcuts.

\textbf{Parameter regression} auxiliary loss enforces continuous structure in the embedding space.
Removing it increases $L_d$ from 1.399 to 1.437 ($+2.7\%$) while improving retrieval R@1 from 67.3 to 70.4, acting as a complementary objective that enriches representations with continuous parameter structure.

\textbf{Dynamic Fx probability scheduling} adaptively adjusts each effect's activation probability based on per-effect loss during training.
Removing it yields the second-largest $L_d$ degradation (1.399 $\to$ 1.465) while again improving retrieval R@1 (67.3 $\to$ 71.7), as upweighting under-learned effects ensures balanced parameter estimation across all effect types.

\textbf{Single-branch} removes the reference branch and encodes only the processed signal, matching the absolute encoding paradigm of Fx-Encoder++~\cite{yeh2025fxencoderpp}.
This leads to retrieval degradation ($-$21.2 R@1, 67.3 $\to$ 46.1) and increases $L_d$ to 1.457 ($+4.1\%$), demonstrating the clear advantage of our dual-branch relative formulation.

Notably, removing cross-segment sampling, parameter regression, or dynamic scheduling \emph{increases R@1} while degrading style transfer. This indicates that retrieval accuracy alone
may not fully reflect style-transfer performance
and is therefore insufficient as the sole metric for effect representations.

\subsection{Bidirectional Encoding Properties}\label{subsec:bidirectional}

Beyond the base model, we additionally train the antisymmetric variant (Eq.~\ref{eq:antisym_diffgate}) under identical configurations. As shown in \tabref{tab:matching} (\emph{Bi-dir.}), it achieves style-transfer performance on par with the base model (Avg $L_d = 1.389$), without sacrificing accuracy under its structural constraint. 
A key empirical property of this variant is that swapping the input order produces an approximately sign-reversed embedding of the reverse-direction relation (Eq.~\ref{eq:bidirectional}).
We evaluate this behavior on 100 triplets per subset from the MUSDB18 drums and vocals subsets using two structural properties:

\begin{enumerate}
    \item \textbf{Antisymmetry}: $\cos(E(A,B), E(B,A)) \!\approx\! {-}1$
    \item \textbf{Composition}: $\cos(E_{AB}\!+\!E_{BC},\, E_{AC}) \!\approx\! 1$
\end{enumerate}

\begin{table}[t]
    \centering
    \caption{Bidirectional encoding properties evaluated on the MUSDB18 drums and vocals subsets. \emph{RelFx (Bi-dir.)} denotes the separately trained antisymmetric variant (Eq.~\ref{eq:antisym_diffgate}), while \emph{RelFx (Base)} denotes the base model (Eq.~\ref{eq:diffgate}).}
    \label{tab:bidirectional}
    \footnotesize
    \begin{tabular}{lccc}
        \toprule
        \textbf{Property} & \textbf{Drums} & \textbf{Vocals} & \textbf{Avg} \\
        \midrule
        \multicolumn{4}{l}{\textit{Antisymmetry} $\cos(E(A,B), E(B,A))$ \; (ideal: $-1.0$)} \\
        \quad RelFx (Bi-dir.) & $\mathbf{-0.998}$ & $\mathbf{-0.998}$ & $\mathbf{-0.998}$ \\
        \quad RelFx (Base)    & $+0.273$ & $+0.281$ & $+0.277$ \\
        \midrule
        \multicolumn{4}{l}{\textit{Composition} $\cos(E(A,B)+E(B,C), E(A,C))$ \; (ideal: $1.0$)} \\
        \quad RelFx (Bi-dir.) & $\mathbf{0.849}$ & $\mathbf{0.834}$ & $\mathbf{0.842}$ \\
        \quad RelFx (Base)    & $0.798$ & $0.742$ & $0.770$ \\
        \bottomrule
    \end{tabular}    
\end{table}

Table~\ref{tab:bidirectional} shows that the sign reversal induced at the fusion layer is nearly preserved in the final normalized embedding ($-0.998$), versus $+0.277$ for RelFx (Base).

\medskip
\noindent\textbf{Limitations.}
RelFx models overall transformations but does not separate source-specific effects within mixtures.
Unseen chain orders and real production chains will be explored in future work.

\section{Conclusion}\label{sec:conclusion}

RelFx is a contrastive framework that learns relative audio-effect transformations rather than absolute effect states and does not require dry references during representation training. It achieves state-of-the-art Fx style-transfer performance under the standard Fx-Encoder++ MUSDB18 protocol across four instrument categories, while its antisymmetric variant supports direction-aware, approximately sign-reversed relation encoding.

\clearpage
\bibliography{ISMIRtemplate}

\section{Ethics Statement}
This work focuses on Fx representation learning for music production applications. The training data consists of licensed music datasets. We do not foresee negative ethical implications of this research.

\end{document}